\documentclass[twocolumn,showpacs,preprintnumbers,amsmath,amssymb]{revtex4}

\usepackage{graphicx}
\usepackage{dcolumn}
\usepackage{bm}

\begin{document}

\preprint{APS/123-QED}

\title{Gilbert damping in non-collinear magnetic systems}

\author{}
\author{S.~Mankovsky, S.~Wimmer, H.~Ebert}
\affiliation{%
Department of Chemistry/Phys. Chemistry, LMU Munich,
Butenandtstrasse 11, D-81377 Munich, Germany 
}%

\date{\today}

\begin{abstract}
The modification of the  magnetization dissipation 
or Gilbert damping
caused by  an inhomogeneous magnetic structure
and expressed in terms of a wave vector dependent 
tensor $\underline{\alpha}(\vec{q})$ is investigated by means of  
linear response theory.
A corresponding  expression for  $\underline{\alpha}(\vec{q})$ 
in terms of the electronic Green function
has been developed giving
in particular the leading 
contributions to the Gilbert damping
linear and quadratic in $q$.
Numerical results for realistic systems are presented
that have been obtained by implementing the scheme 
within the framework  
of  the  fully relativistic KKR (Korringa-Kohn-Rostoker)
band structure method.
Using the multilayered system (Cu/Fe$_{1-x}$Co$_x$/Pt)$_n$
as an example for systems without inversion symmetry we
 demonstrate the occurrence of  non-vanishing linear contributions.
 For the alloy system  bcc Fe$_{1-x}$Co$_x$ having inversion symmetry,
 on the other hand, only the quadratic contribution is non-zero.
 As it is shown, this quadratic contribution does not vanish even
 if the spin-orbit coupling is suppressed, i.e.\ it is a direct
 consequence of the non-collinear spin configuration. 
\end{abstract}

\maketitle

\section{ Introduction \label{IN}}

The magnetization dissipation in magnetic materials is conventionally 
characterized by means of the Gilbert damping (GD) tensor 
$\underline{\alpha}$
that enters the Landau-Lifshitz-Gilbert  (LLG) equation \cite{Gil04}.
This positive-definite second-rank tensor depends in
general  on the magnetization direction. 
It is well established that in the case of spatially uniformly magnetized 
ferromagnetic (FM) metals two regimes of slow  magnetization dynamics can be distinguished,
 which are governed by different mechanisms of dissipation  \cite{Kam70,FS06,GIS07}:
a conductivity-like behaviour occuring in the limiting case of
 ordered compounds that may be connected to the 
 Fermi breathing mechanism and a
resistivity-like behaviour shown by  materials with
appreciable structural, chemical or temperature induced disorder
and connected to a spin-flip scattering mechanism.
An important  issue  is that both mechanisms are 
determined by the spin-orbit coupling
 in the system (see e.g. \cite{MKWE13,Kam70,GIS07}). 
 During the last years, it
 was demonstrated by various authors  that 
  first-principles calculations for the GD parameter
for collinear ferromagntic materials  allow to cover both regimes 
without use of any phenomenological
 parameters. In fact, in spite of the differences 
 concerning the formulation for the damping parameter and 
 the corresponding  implementaion
 \cite{SKB+10,EMKK11, TKD15}, 
 the numerical  results are in general in rather  good agreement with each other  as well as with experiment.

In the case of a pronounced non-collinear magnetic texture, e.g.\ in
the case of  domain walls or 
topologically nontrivial magnetic configurations like skyrmions, 
the description of the magnetization dissipation 
assuming a spatial-invariant tensor $\underline{\alpha}$ is incomplete, and a non-local character of GD
tensor in such systems has to be taken 
into account \cite{THV09, HFR15, HB14}.
This implies that the dissipative torque on the magnetization 
should be represented by the expression of the following general form
 \cite{UMS15}:
\begin{eqnarray}
\label{GD_NL_torque}
\tau_{GD} &=& \hat{m}(\vec{r},t) \times \int d^3r'
\underline{\alpha}(\vec{r} - \vec{r}\,')
\, \frac{\partial}{\partial t}
\hat{m}(\vec{r}\,',t) \;.
\end{eqnarray}
%
In the case of a magnetic texture varying slowly in space, 
however, an 
expansion of the damping parameter 
in terms of the magnetization density and its gradients \cite{HB14}
is nevertheless appropriate:
%
\begin{eqnarray}
\label{GD_NL_torque_Tailor}
\alpha_{ij} &=& \alpha_{ij} + \alpha_{ij}^{kl}m_km_l
+\alpha_{ij}^{klp}m_k\frac{\partial}{\partial r_l}m_p \\
&& +
\alpha_{ij}^{klpq}\frac{\partial}{\partial
  r_k}m_l\frac{\partial}{\partial r_p}m_q + ...  \; ,  \nonumber
\end{eqnarray}
where the first term $ \alpha_{ij}$ stands for the conventional 
isotropic GD  and the second
term $\alpha_{ij}^{kl}m_km_l$  is associated with the magneto-crystalline
anisotropy (MCA). 
The third so-called chiral term
$\alpha_{ij}^{klp}m_k\frac{\partial}{\partial r_l}m_p $ 
is non-vanishing in non-centrosymmetric systems. 
The important role of this contribution to the 
damping was demonstrated experimentally 
when investigating the field-driven
domain wall (DW) motion in asymmetric Pt/Co/Pt trilayers \cite{JSD+15}.

As an alternative to the expansion in Eq.\ (\ref{GD_NL_torque_Tailor})
one can discuss the Fourier transform $\underline{\alpha}(\vec{q})$
of the damping parameter characterizing inhomogeneous magnetic systems,
which enter the spin dynamics equation 
\begin{eqnarray}
\label{LLG_q}
\frac{\partial }{\partial t}\vec{m}({\vec{q}}) = -\gamma
\vec{m}({\vec{q}}) \times \vec{H} - \vec{m}({\vec{q}}) \times
\underline{\alpha}({\vec{q}})\frac{\partial }{\partial
  t}\vec{m}({\vec{q}}) \;.
\end{eqnarray}
%
In this formulation  the  term linear in $q$ 
 is the first chiral term appearing in
the  expansion of  $\underline{\alpha}(\vec{q})$ in powers of $q$. 
Furthermore, it  is  important to note that it is directly
connected to the $\alpha_{ij}^{klp}m_k\frac{\partial}{\partial
  r_l}m_p$  term in Eq.\ (\ref{GD_NL_torque_Tailor}).

By applying a gauge field theory, the origin of the non-collinear corrections to the GD 
can be ascribed to the  emergent electromagnetic field
created in the time-dependent magnetic texture \cite{ZZ09,TN14}.  
Such an emergent electromagnetic 
field gives rise to a spin current
whose divergence characterizes the change of the angular momentum in
the system. This allows to discuss the impact of
 non-collinearity on the 
 GD via a  spin-pumping formulation \cite{TM08,ZZ09,THV09}. 
Some details of the physics behind this effect depend on the
specific properties of the  material considered. 
Accordingly, different models for magnetisation dissipation were
discussed in the literature \cite{THV09,ZZ09,ZMHN11,UMS15,CBT15,AMGM16}.
Non-centrosymmetric two-dimensional systems for 
 which the Rashba-like spin-orbit coupling plays
an important role have received special interest in this context. 
They have been discussed in particular by Akosa et
al.\ \cite{AMGM16}, in order to explain the origin of chiral
GD in the presence of a chiral magnetic structure.

The fourth  term on the r.h.s.\ of  Eq.\ (\ref{GD_NL_torque_Tailor})
corresponds to a quadratic term  of an expansion of  $\underline{\alpha}(\vec{q})$
with respect to $q$.
It was investigated for bulk systems with non-magnetic \cite{HVT08}
and magnetic \cite{THV09} impurity atoms, 
for which the authors have shown on the basis of model consideration
that it can give a significant correction to the 
homogeneous GD in the
case of weak metallic ferromagnets. 
In striking contrast to the  uniform part  of the GD  
this contribution does not require a non-vanishing  spin-orbit
interaction. 

To  our knowledge, only very few
 ab-initio investigations 
on the  Gilbert damping in non-collinear 
magnetic systems along the lines sketched above 
have been  reported so far in the literature. 
Yuan et al.\ \cite{YHL+14} calculated the in-plane 
and out-of-plane damping
parameters in terms of the scattering matrix for permalloy in the
presence of N\'{e}el and Bloch domain walls.
Freimuth et al. \cite{FBM17a}, discuss the properties 
of a $q$-dependent Gilbert damping $\underline{\alpha}(\vec{q})$
 calculated for the one-dimensional Rashba model
in the presence of the N\'{e}el-type non-collinear magnetic exchange
field, demonstrating different GD 
for  left-handed and right-handed DWs.
Here we extend the formalism developed before  to deal with the GD in
ferromagnets  \cite{SKB+10}, to get access to non-collinear system. The formalism based on linear
response theory allows to expand the GD parameters with respect to a
modulation of the magnetization expressed in terms of a wave vector
$\vec{q}$. Corresponding numerical results will be presented and
discussed.

 \section {\bf Gilbert damping for non-collinear magnetization }

In the following we focus on  the intrinsic contribution to
the  Gilbert damping, 
excluding spin current induced magnetization dissipation which 
occurs in the presence of an external electric field.
For the considerations on the  magnetization dissipation
an  adiabatic variation of the magnetization 
in the  time and space domain is assumed. 
Moreover, it is assumed that the magnitude of 
the local magnetic moments is unchanged 
during a change of the  magnetization, 
i.e.\ the exchange field should be strong enough 
to separate transverse and longitudinal parts
of the  magnetic susceptibility. 
With these restrictions, the {\em non-local}
 Gilbert damping can be determined
in terms of the spin susceptibility tensor 
\begin{eqnarray} 
\chi_{\alpha\beta}(\vec{q},\omega) & = & 
i \frac{1}{V}\int\limits_0^{\infty}dt
\langle {\hat S}_\alpha (\vec{q}, t)  {\hat S}_\alpha (-\vec{q}, 0) \rangle_0e^{i(\omega - \delta)t} \;,
\label{PI_TORQUE}
\end{eqnarray}
where ${\hat S}_\alpha (\vec{q}, t) $
is the $\vec{q}$- and  $t$-dependent spin operator and 
 reduced units have been used ($\hbar = 1$).
With this, the Fourier transformation of the real-space Gilbert damping
can be  represented by the expression \cite{QV02, HVT07}
\begin{eqnarray}
  \alpha_{\alpha\beta}(\vec{q}) &=&  
  \frac{\gamma}{M_0V} \lim_{\omega \to 0} \frac{\partial  \Im
[\chi^{-1}]_{\alpha\beta}(\vec{q}, \omega)}{\partial \omega} \;.
\label{GD_HTV08}
\end{eqnarray}
%
Here $\gamma = g \mu_B$ is the gyromagnetic ratio, $M_0 =
\mu_{tot} \mu_B / V$ is the equilibrium
magnetization and $V$ is the volume of the system.
In order to avoid the calculation of the
dynamical magnetic susceptibility tensor 
$\underline{{\chi}}(\vec{q}, \omega)$,
which is the Fourier transformed of
the real space susceptibility 
 $\underline{{\chi}}(\vec{r}- \vec{r}\,',\omega)$,
 it is convenient to represent 
  $\underline{\chi}(\vec{q},\omega) $
in Eq.\ (\ref{GD_HTV08}), 
in terms of a correlation function of time derivatives 
of $\hat S$.
As  $\dot{\hat S}$ corresponds to the  torque
$\vec{\cal{T}}$, that may  include non-dissipative and dissipative parts,
one may consider instead the torque-torque correlation function
 $\underline{\pi}(\vec{q}, \omega)$
\cite{MK62,Mah70a,Has71,HVT07}.

Assuming the
magnetization direction parallel to $\hat z$
one obtains the expression for the Gilbert damping $\underline{\alpha}({\vec{q}})$ 
\begin{eqnarray}
\underline{\alpha}({\vec{q}})  =   \frac{\gamma}{M_0V} \lim_{\omega \to 0}
\frac{\partial  \Im[ \underline{\epsilon} \cdot
  \underline{\pi}(\vec{q}, \omega) \cdot
  \underline{\epsilon}]}{\partial \omega} \; .
\label{GD_Torque}
\end{eqnarray}
%
where 
 $ \underline{\epsilon}  = \begin{bmatrix}
   0 & 1 \\
   -1 & 0     
 \end{bmatrix} $
is the transverse Levi-Civita tensor.
This implies the following relationship of the
$\underline{\alpha}$ tensor elements 
with the elements of the torque-torque correlation 
tensor $\underline{\pi}$: $\alpha_{xx} \sim - \pi_{yy}$ and
$\alpha_{yy} \sim -\pi_{xx}$ \cite{HVT07}.

Using  Kubo's linear response theory 
in the Matsubara representation
and taking into account the translational symmetry of a solid
the torque-torque correlation function $\pi_{\alpha\beta}(\vec{q}, \omega)$
can be expressed by  (see, e.g. \cite{Mah00}):
%
\begin{eqnarray}
\underline{\pi}_{\alpha\beta}(\vec{q}, i \omega_n) &=& \frac{1}{\beta} \sum_{p_m}
\langle {\cal T}^{\alpha} {\cal G}(\vec{k} + \vec{q}, i\omega_n + ip_m) 
\nonumber \\
&& \qquad \qquad
{\cal T}^{\beta} {\cal G}(\vec{k}, ip_m) \rangle_c \; ,
\label{CHI_Mazu}
\end{eqnarray}
where ${\cal G}(\vec{k}, ip)$ is the Matsubara Green function
and 
$\langle ... \rangle_c$ indicates a configurational average 
required in the presence of any disorder (chemical, structural 
or magnetic) in the system. 
Using a Lehman representation for the  Green function \cite{Mah00} 
\begin{eqnarray}
{\cal G}(\vec{k}, ip_m) &=& \int_{-\infty}^{+\infty}
\frac{dE}{\pi} \frac{\Im {G}^+ (\vec{k}, E)}{ip_m - E} 
\label{GF_Mazu2}
\end{eqnarray}
with $G^+ (\vec{k}, E)$ the retarded Green function and
  using the relation
\begin{eqnarray*}
 \frac{1}{\beta} \sum_{p_m} \frac{1}{ip_m + i\omega_n -
  E_1}\frac{1}{ip_m - E_2} &=& \frac{f(E_2) - f(E_1)}{i\omega_n + E_2 -  E_1}
\label{GF_Mazu4}
\end{eqnarray*}
for the sum over  the Matsubara poles in Eq.\ (\ref{CHI_Mazu}),
the torque-torque correlation function
is obtained as:
\begin{widetext}
\begin{eqnarray}
\underline{\pi}_{\alpha\beta}(\vec{q}, i \omega_n) &=&
\frac{1}{\Omega_{BZ}} \int d^3k  \int\limits_{-\infty}^{+\infty} \frac{d
  E_1}{\pi}  \int\limits_{-\infty}^{+\infty} \frac{d E_2}{\pi}  {\mbox
  {Tr}}  \bigg\langle {\cal T}^{\alpha} \Im {G}(\vec{k}, E_1) {\cal
  T}^{\beta} \Im {G} (\vec{k}, E_2) \frac{f(E_2) - f(E_1)}{i\omega_n + E_2 -  E_1} \bigg\rangle_c \;.
\label{GF_Mazu3}
\end{eqnarray}
%
Perfoming finally the analytical continuation 
$i\omega_n \to \omega + i\delta$
one arrives at the   expression 
%
\begin{eqnarray}
\Gamma_{\alpha\beta}(\vec{q}, \omega) &=& -   
\frac{\pi}{\Omega_{BZ}}  \int d^3k \int\limits_{-\infty}^{+\infty} \frac{d E_1}{\pi}  \int\limits_{-\infty}^{+\infty} \frac{d E_2}{\pi} {\mbox {Tr}} \bigg\langle {\cal T}^{\alpha} \Im {G}(\vec{k} + \vec{q}, E_1)
 {\cal T}^{\beta} \Im {G}(\vec{k}, E_2) \bigg\rangle_c (f(E_2) - f(E_1)) \delta (\omega + E_2 -  E_1) \nonumber \\
 &=& -  
 \frac{\pi}{\Omega_{BZ}} \int d^3k \int\limits_{-\infty}^{+\infty}
 \frac{d E}{\pi}  {\mbox {Tr}}  \bigg\langle {\cal T}^{\alpha} \Im {G} (\vec{k} + \vec{q}, E)
  {\cal T}^{\beta} \Im {G}(\vec{k}, E + \omega)\bigg\rangle_c (f(E) - f(E + \omega))
\label{GF_Mazu5}
\end{eqnarray}
 for the imaginary part of
the correlation
function with $\Gamma_{\alpha\beta}(\vec{q}, \omega)  = -\pi \Im
\pi_{\alpha\beta}(\vec{q}, \omega)$.
Accordingly one gets for the diagonal elements of
 Gilbert damping tensor the expression
%
\begin{eqnarray}
\alpha_{\alpha\alpha}(\vec{q}) &=&  \frac{\gamma}{M_0V} \lim_{\omega \to 0}
\frac{\partial [\underline{\epsilon} \cdot
  \underline{\Gamma}(\vec{q}, \omega) \cdot
  \underline{\epsilon}]}{\partial \omega} \Bigg|_{\alpha\alpha} 
  \nonumber \\
& = &  \frac{\gamma\pi}{M_0V} \lim_{\omega \to 0} \frac{\partial}{\partial
  \omega}\frac{1}{\Omega_{BZ}} \int d^3k 
\int\limits_{-\infty}^{+\infty} \frac{d E}{\pi^2} (f(E + \omega) - f(E))
 {\mbox {Tr}} \bigg\langle {\cal T}^{\beta} \Im {G}(\vec{k} + \vec{q}, E)
  {\cal T}^{\beta} \Im {G}(\vec{k}, E + \omega)\bigg\rangle_c 
  \nonumber \\
& = &  \frac{\gamma}{M_0V} 
\frac{1}{\Omega_{BZ}} \int d^3k \int\limits_{-\infty}^{+\infty} \frac{d E}{\pi} \delta(E - E_F)
  {\mbox {Tr}} \bigg\langle {\cal T}^{\beta}  \Im G(\vec{k} + \vec{q}, E)
  {\cal T}^{\beta} \Im G(\vec{k}, E)\bigg\rangle_c 
  \nonumber \\
& = & \frac{1}{4}[\alpha_{\alpha\alpha}(\vec{q},G^+,G^+) + \alpha_{\alpha\alpha}(\vec{q},G^-,G^-) - 
\alpha_{\alpha\alpha}(\vec{q},G^+,G^-) - \alpha_{\alpha\alpha}(\vec{q},G^-,G^+) ]
\;,
\label{GD_CHI}
\end{eqnarray}
where the index $\beta$ of the torque operator ${\cal T}^{\beta}$
is related to the index $\alpha$ according to Eq.\ \ref{GD_Torque}, and
the auxiliary functions
\begin{eqnarray}
\alpha_{\alpha\alpha}(\vec{q},G^\pm,G^\pm) &=& 
\frac{\gamma}{M_0V\pi} 
\frac{1}{\Omega_{BZ}} \int d^3k  {\mbox {Tr}} \bigg\langle
{\cal T}^{\beta} G^\pm(\vec{k} + \vec{q}, E_F){\cal T}^{\beta} G^\pm(\vec{k}, E_F)  \bigg\rangle_c
\label{GD_CHI-aux}
\;
\end{eqnarray}
expressed in terms of the retarded and advanced Green functions,
$G^+$ and $G^-$, respectively.
\medskip

To account properly for the impact of spin-orbit coupling
when  dealing with Eqs.\ (\ref{GD_CHI}) and (\ref{GD_CHI-aux}) 
a description of the electronic structure based on the 
fully relativistic Dirac formalism is used. 
Working within the framework of local
spin density  formalism (LSDA) this implies for the 
Hamiltonian the form  \cite{Ebe00}: 
%
\begin{eqnarray}
\hat{H}_D = c\vec{\bm\alpha}\cdot\vec{p} +\beta  m c^2 +
 V(\vec r) + \beta \vec{\bm\sigma}\cdot\hat{\vec m}B_{xc}(\vec r)
\; .
\label{Hamiltonian}           
\end{eqnarray}
%
Here $\alpha_i$ and $\beta$  are the standard Dirac
matrices,
$\vec{\bm\sigma}$ denotes the vector of relativistic Pauli matrices,
$\vec p $  is the relativistic momentum operator \cite{Ros61} and
the functions $V(\vec r)$ and
$\vec B_{xc} = \vec{\bm\sigma}\cdot\hat{\vec m}B_{xc}(\vec r)$  are the
spin-averaged and spin-dependent  parts, respectively, 
of the LSDA potential \cite{MV79} with $\hat{\vec m}$
giving the orientation of the magnetisation.

With the Dirac Hamiltonian given by Eq.\ (\ref{Hamiltonian}),
the torque operator 
may be
written as
 $ \vec{\cal{T}} = \beta [\vec{\sigma} \times \hat{\vec m}] B_{xc}(\vec r)$. 
Furthermore, the Green functions entering Eqs.\ (\ref{GD_CHI}) and (\ref{GD_CHI-aux})
are determined using the spin-polarized relativistic version of multiple
 scattering theory \cite{Ebe00,EBKM16}
 with the real space representation of the retarded Green function
 given  by:
%
\begin{eqnarray}
 G^{+}(\vec{r},\vec r\,',E)& = &
 \sum_{\Lambda \Lambda'}
Z^{n}_{\Lambda}(\vec r,E)
\,
  {\tau}_{ \Lambda  \Lambda'}^{nm}(E)
\,
Z^{m\times}_{\Lambda'}(\vec r\,',E)
\nonumber \\
&&
\hspace{-9.3ex}
-
\delta_{nm}
\sum_{\Lambda }
\left[
Z^{n}_{\Lambda}(\vec r,E)
\,
J^{n\times}_{\Lambda'}(\vec r',E)
\Theta(r_n'-r_n)
\right.
\nonumber
\\
&&
\left.
  +
J^{n}_{\Lambda}(\vec r,E)
\,
Z^{n\times}_{\Lambda'}(\vec r\,',E)
\Theta(r_n-r_n')
\right]
.
\label{GreensFunction}
\end{eqnarray}
%
Here $\vec{r},\vec{r}\,'$ refer to atomic cells centered at sites
 $n$ and $m$, respectively, where
$Z^{n}_{\Lambda}(\vec r,E)=Z_{\Lambda}(\vec r_n,E) 
= Z_{\Lambda}(\vec r-\vec R_n ,E) $ 
is a function centered at the corresponding lattice vector $\vec R_n$.
The four-component wave functions $Z^{n}_{\Lambda}(\vec r,E)$ 
($J^{n}_{\Lambda}(\vec r,E)$) are
regular (irregular)
solutions to the single-site Dirac equation labeled by the
combined quantum numbers $\Lambda = (\kappa,\mu)$, with
$\kappa$ and $\mu$  being the spin-orbit and magnetic quantum numbers
\cite{Ros61}.
Finally, ${\tau}^{nm}_{ \Lambda  \Lambda'} (E)$ is
 the so-called scattering path operator
 that transfers an electronic wave coming in at 
site $ m $ into a wave going out from site $ n $ with
all possible intermediate scattering events accounted for. 

 Using matrix notation with respect to $\Lambda$, this
leads to the following expression for the auxilary damping parameters
in Eq.\ (\ref{GD_CHI-aux}):
\begin{eqnarray}
\alpha_{\alpha\alpha}(\vec{q},G^\pm,G^\pm) &=&  \frac{\gamma}{M_0V\pi} 
\frac{1}{\Omega_{BZ}} \int d^3k {\mbox {Tr}} \bigg\langle
{\underline T}^{\beta} {\underline \tau}(\vec{k} + \vec{q}, E_F^\pm){\underline
  T}^{\beta} {\underline \tau}(\vec{k}, E_F^\pm) \bigg\rangle_c \;.
\label{GD_CHI-aux2}
\end{eqnarray}
In the case of a  uniform magnetization, i.e.\ for  $q = 0$
one obviously gets an 
expression for the Gilbert damping tensor 
as it was worked out before \cite{EMKK11}.
Assuming small wave vectors, the 
term ${\underline \tau}(\vec{k} + \vec{q}, E_F^\pm$)
 can be expanded w.r.t.\ $\vec{q}$ leading to the series
%
\begin{eqnarray}
{\underline \tau}(\vec{k} + \vec{q}, E_F)  & = &  {\underline \tau}(\vec{k}, E) + \sum_\mu \frac{\partial
   {\underline \tau}(\vec{k}, E)}{\partial k_\mu} q_\alpha + \frac{1}{2} \sum_{\mu\nu} \frac{\partial
  {\underline \tau}(\vec{k}, E)}{\partial k_\mu \partial k_\nu }  q_\mu q_\nu
 + ...
\label{tau-expansion}
\end{eqnarray}
that results in a corresponding  expansion for the Gilbert damping: 
\begin{eqnarray}
{\underline \alpha}(\vec{q})  & = &  {\underline \alpha} 
+ \sum_\mu  \underline {\alpha}^\mu q_\mu 
+ \frac{1}{2} \sum_{\mu\nu} \underline {\alpha}^{\mu\nu}   q_\mu q_\nu
 + ...
\label{GD_CHI_Q}
\end{eqnarray}
%
with the following expansion coefficients:
%
\begin{eqnarray}
\label{GD_CHI0}
\alpha^{0 \pm\pm}_{\alpha\alpha} & =  &  \frac{g}{\pi \mu_{tot}}
\frac{1}{\Omega_{BZ}} {\mbox{Trace}} \int d^3k  \bigg\langle
 \underline{T}_{\beta}   \underline{\tau}(\vec{k}, E_F^{\pm})
 \underline{T}_{\beta}   
 \underline{\tau}(\vec{k}, E_F^{\pm})  \bigg\rangle_c
\\
%
\alpha_{\alpha\alpha}^{\mu \pm\pm} & = &  \frac{g}{\pi \mu_{tot}} 
\label{GD_CHI1}
\frac{1}{\Omega_{BZ}} {\mbox{Trace}} \int d^3k  \bigg\langle
 \underline{T}^{\beta}    \frac{\partial
  \underline{\tau}(\vec{k}, E_F^{\pm})}{\partial k_\mu} 
  \underline{T}^{\beta}    
  \underline{\tau}(\vec{k}, E_F^{\pm}) \bigg\rangle_c
  \\
\alpha_{\alpha\alpha}^{\mu\nu \pm\pm} & = &   \frac{g}{\pi \mu_{tot}} 
\label{GD_CHI2}
\frac{1}{2\Omega_{BZ}}{\mbox{Trace}}  \int d^3k \bigg\langle 
\underline{T}^{\beta}   \frac{\partial^2
  \underline{\tau}(\vec{k}, E_F^{\pm})}{\partial k_\mu \partial k_\nu} 
  \underline{T}^{\beta}    \underline{\tau}(\vec{k}, E_F^{\pm}) \bigg\rangle_c
\; ,
\end{eqnarray}
%
and with the g-factor $2(1 + {\mu_{orb}}/{\mu_{spin}})$ in terms of
the spin and orbital moments, $\mu_{spin}$ and $\mu_{orb}$,
respectively, and  the total magnetic moment $\mu_{tot} =
\mu_{spin}+\mu_{orb}$. 
The numerically cumbersome term in Eq.\ (\ref{GD_CHI2}),
that involves the second order
derivative of the matrix of $\vec{k}$-dependent
 scattering path operator $\tau(\vec{k},E)$, 
 can be reformulated by means
of an  integration by parts:
\begin{eqnarray*}
\frac{1}{\Omega_{BZ}}\int d^3k \underline{T}_\beta \; \underline{\tau}(\vec{k},E_F)\;
\underline{T}_\beta\;  \frac{\partial^2
  \underline{\tau}(\vec{k},E_F) }{\partial k_\mu \partial
  k_\nu} 
&=& \Bigg[ \underbrace{  \int\int dk_\beta  dk_\gamma \underline{T}_\beta^i \; \underline{\tau}(\vec{k},E)\;
\underline{T}_\beta^j\; \frac{\partial
  \underline{\tau}(\vec{k},E) }{\partial
  k_\beta} \Bigg|_{-\frac{K_{\alpha}}{2}}^{\frac{K_{\alpha}}{2}}\;}_{=0}  \\
&& -  \int\int\int dk_\alpha  dk_\beta  dk_\gamma \underline{T}_\beta \; \frac{\partial
  \underline{\tau} (\vec{k},E_F)}{\partial k_\mu} \;
\underline{T}_\beta\; \frac{\partial \underline{\tau}(\vec{k},E_F) }{\partial
  k_\nu}\;\Bigg]   \\
&=& - \frac{1}{\Omega_{BZ}}  \int d^3k \underline{T}_\beta \; \frac{\partial
  \underline{\tau} (\vec{k},E_F)}{\partial k_\mu} \;
\underline{T}_\beta\; \frac{\partial \underline{\tau}(\vec{k},E_F) }{\partial
  k_\nu}\;
\label{Denergy1}
\end{eqnarray*}
leading to the much more convenient expression:
\begin{eqnarray}
\label{GD_CHI3}
\alpha_{\alpha\alpha}^{\mu\nu \pm\pm} & = &
  - \frac{g}{2\pi\mu_{tot}} 
\int d^3k  {\mbox{Tr}} \bigg\langle \underline{T}^{\beta}   \frac{\partial
  \underline{\tau}(\vec{k}, E_F^\pm)}{\partial k_\mu} 
  \underline{T}^{\beta}  \frac{\partial
   \underline{\tau}(\vec{k}, E_F^\pm)}{\partial k_\nu} \bigg\rangle_c
   \; .
\end{eqnarray}

\end{widetext}

\section{Results and discussions}

\begin{figure}[h]
\includegraphics[width=0.45\textwidth,angle=0,clip]{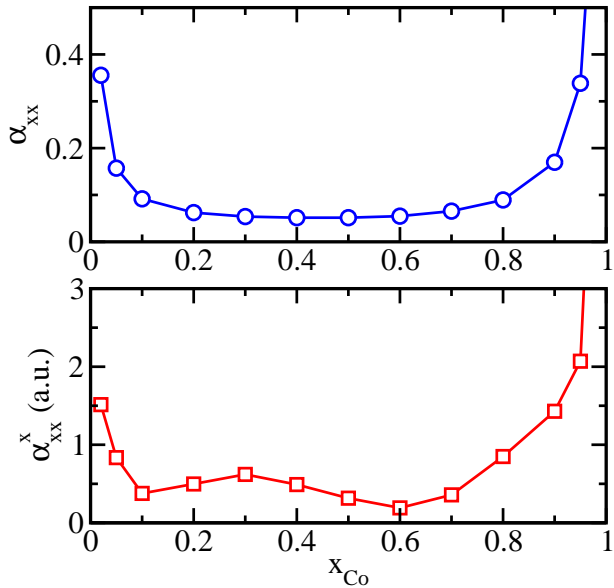}\;
\caption{\label{fig:GD1} The  Gilbert damping parameters 
$\alpha_{xx}$ (top) and $\alpha^x_{xx}$  (bottom) calculated 
for the model multilayer system (Cu/Fe$_{1-x}$Co$_x$/Pt)$_n$  using 
 Eqs.\ (\ref{GD_CHI0}) and (\ref{GD_CHI1}), respectively.}    
\end{figure}

The scheme presented above to deal with the Gilbert damping in non-collinear systems
has been implemented 
within the SPR-KKR program package \cite{SPR-KKR7.7}.
To examine the importance of the chiral correction to
the Gilbert damping a first application of 
  Eq.\ (\ref{GD_CHI1}) has been made  for the  
multilayer system (Cu/Fe$_{1-x}$Co$_x$/Pt)$_n$ seen as a 
non-centrosymmetric model system.
The calculated  zero-order (uniform) GD parameter
 $\alpha_{xx}$ and the corresponding 
first-order (chiral) $\alpha^{x}_{xx}$ correction
 term  for $\vec{q}\, \| \hat{x}$ are plotted in
Fig.\ \ref{fig:GD1} top  and bottom, respectively,
as a function of the Fe concentration $x$. 
%
Both terms,
$\alpha_{xx}$  and $\alpha^x_{xx}$, increase approaching
the pure limits w.r.t.\ the  Fe$_{1-x}$Co$_x$ alloy subsystem. 
In the case of the  uniform parameter  $\alpha_{xx}$, 
this increase is associated with the dominating
 breathing Fermi-surface damping mechanism. 
 This  implies  that 
the modification of the Fermi surface (FS) 
induced by the spin-orbit coupling (SOC)
follows the
magnetization direction that slowly varies with time. 
An additional contribution to the GD, having a similar origin,
occurs for the non-centrosymmertic systems with helimagnetic structure. 
In this case, the features of the electronic structure governed by the
lack of inversion symmetry result in a FS modification dependent on the
helicity of the magnetic structure.
This implies a chiral contribution to the GD which
can be associated with the term proportional to the gradient of the
magnetization. 
Obviously, this additional modification of the FS and the associated
mechanism for the GD does not show up for a uniform ferromagnet.
As $\underline\alpha$ is caused by the SOC one can expect that it vanishes
for vanishing SOC. This was indeed demonstrated before \cite{MKWE13}.
The same holds also for  $\underline\alpha^x$ that is cased by SOC as well.
 \medskip

\begin{figure}[h]
\includegraphics[width=0.45\textwidth,angle=0,clip]{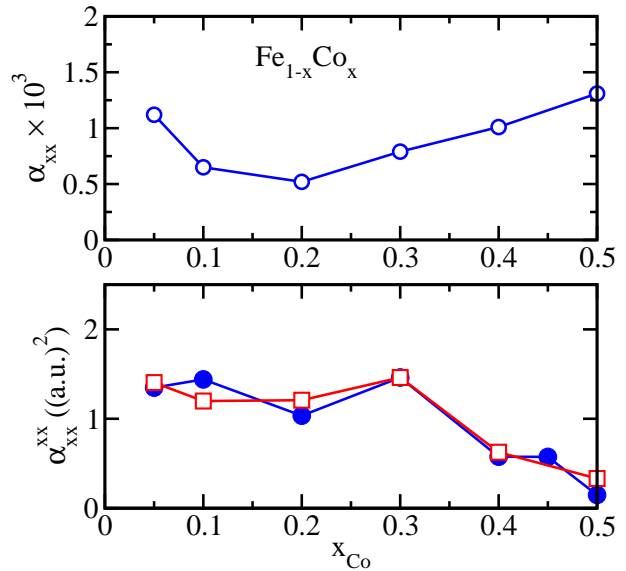}\;
\caption{\label{fig:GD2} The  Gilbert damping 
terms $\alpha_{xx}$ (top) and $\alpha^{xx}_{xx}$ (bottom) 
calculated for bcc Fe$_{1-x}$Co$_x$.}   
\end{figure}

Another system considered is the ferromagnetic alloy system 
 bcc Fe$_{1-x}$Co$_x$.  
As this system has inversion symmetry the first-order
term  $\underline\alpha^\mu$ should vanish. 
This expectation could also be confirmed by 
calculations that account for the SOC.
The next non-vanishing term of  the  expansion of the GD
 is the term $\propto q^2$.
The corresponding second-order term $\alpha^{xx}_{xx}$ is plotted in 
Fig. \ref{fig:GD2} (bottom) together with the 
zero-order term
$\alpha_{xx}$ (top). 
The bottom panel shows in addition results for 
 $\alpha^{xx}_{xx}$ that have been 
obtained by calculations with the SOC suppressed.
As one notes the results for the full SOC and for SOC suppressed
are very close to each other.
The small difference between the curves for that reason 
have to be ascribed to the hybridization of the spin-up
and spin-down subsystems due to SOC.
As discussed in the literature \cite{HVT08, THV09,ZMHN11}
a non-collinear magnetic texture has a corresponding consequence
but a much stronger impact here.
 In contrast to the GD in uniform FM systems
where SOC is required to break the total spin conservation in the system, 
$\alpha^{xx}_{xx}$ is associated with the spin-pumping effect
that can be ascribed to an emergent electric field 
created in the non-uniform magnetic system.
In this case magnetic dissipation occurs due to the
  misalignment of the electron spin following the dynamic magnetic profile
  and the magnetization orientation at each atomic site, leading to 
  the dephasing of electron spins \cite{TM08}

\section{Summary}

To summarize, expressions for corrections to the GD of 
homogeneous systems  were derived  which are
expected to contribute in the case of non-collinear magnetic systems. 
The expression for the  GD parameter $\alpha(\vec{q})$ seen as
a function of the wave vector $\vec{q}$ is
expanded in powers of $q$.
In the limit of weakly varying magnetic textures, 
this leads to the standard
uniform term, $\underline\alpha$, 
and the  first- and second-order
corrections, $\underline\alpha^\mu$ and $\underline\alpha^{\mu\nu}$,
respectively.
Model calculations confirmed  that a non-vanishing 
value for $\underline\alpha^\mu$
can be expected for systems without inversion symmetry.
In addition, SOC has been identified as the major source for this term.
The second-order term, on the other hand, may also show up
for systems with inversion symmetry. In this case it was
demonstrated by numerical work, that 
SOC plays only a minor role for $\underline\alpha^{\mu\nu}$,
while the non-collinearity of the magnetization plays the central role.

\section{Acknowledgement}

Financial support by the DFG via SFB 1277 (Emergente relativistische Effekte in der Kondensierten Materie) is gratefully acknowledged.


\end{document}